\documentclass[12pt]{article}
\usepackage{amsfonts}
\begin{document}
\title{Inverse eigenvalue problem
for discrete three-diagonal Schr\"odinger operator and the
continuum limit in Tikhonov regularization approach. }

\author{V. M. Chabanov \\
Space Research Institute, \\
Profsoyuznaya street, 84/32, Moscow, 117997, Russia \\
 } \date{}

\maketitle

\begin{abstract}
In the present article  the self-contained derivation  of main
formulas of the inverse eigenvalue problem for the Sturm-Liouville operator  
is given on the base of its discrete approximation and the subsequent transition 
to the limit of the continuum coordinate. In doing so, 
we adhere to Tikhonov concept, 
where the discrete approximation is performed as a regularization of the ill-posed operator problem.  
While the continuum limit for the three-diagonal matrix Hamiltonian is impossible as such, 
following the Tikhonov procedure leads not to a limit for the potential matrix itself, but rather for 
a certain combination of the matrix elements, and this limit is precisely the desired 
potential in the continuum case.

\end{abstract}

\section{Introduction}

Without any doubt, the inverse problem theory for the Schrödinger operator [1-6] is one 
of the most beautiful achievements of mathematical physics. A striking demonstration 
of the capabilities of this method is the derivation of exact solutions to nonlinear equations 
\cite{AblSeg}. It is also widely used to reconstruct potentials from scattering phases for 
one-dimensional problems \cite{Geramb}. A fundamental limitation is the overdetermination 
of scattering data in multidimensional ($D \ge 2$) problems, which leads to the reconstruction 
of nonlocal potentials, and only special constraints permit remaining within the class of local 
potentials \cite{Fad}. The corresponding equations for such constraints are too complicated 
for numerical computation. Nevertheless, even here there are fairly unexpected possibilities 
associated, for example, with the fact that the inverse problem equations admit the broadest 
class of explicit analytical solutions — the so-called Bargmann-type potentials \cite{Barg}, or 
exactly solvable inverse problem models. This makes it possible to construct quantum systems 
with predetermined spectral properties using simple analytical expressions — shifting energy 
levels, modifying the reduced widths of resonances, and obtaining quantum systems with entirely 
unique properties; see [11-13]. Although this has no direct practical application, it greatly helps 
deepen our understanding of the very nature of the quantum world.

As is often the case, the development of the inverse problem for the Sturm–Liouville operator proceeded 
largely independently and asynchronously along two directions: the continuous problem and the 
discrete-variable problem, or lattice problems. The latter are obviously of fundamental importance as a 
means of numerically solving inverse problem equations and connecting these two problems is  a very important task. 
This means that we must derive the previously known inverse problem 
equations as the continuum limit when the discrete-variable step tends to zero. Such studies have indeed 
been carried out, and one work that at least influenced the author personally is [14-16], where the case of a 
local single-diagonal finite-difference (without additional bands) matrix potential was considered. One should 
also mention the work \cite{Berry}, where a potential with neighboring bands in the interaction matrix was 
studied for the case of a Hamiltonian with a continuous spectrum. We also considered such a case in \cite{Ch}. 
There, all inverse problem equations for the Schrödinger operator were derived in the Gelfand–Levitan formalism 
for a finite interval using neighboring potential bands. However, in that work we limited ourselves to the case 
where the very existence of convergence of the discrete inverse problem to the continuum limit was assumed 
a priori. This cannot serve as a sufficient and complete derivation, although it allowed us to discern the main 
outlines of the solution to the problem of the transition from the discrete to the continuous variable. The difficulty 
of such a problem is evident if only because, as is well known, the solution of the discrete inverse problem (the matrix 
discrete potential) has no continuum limit.

After some time we returned to this question, but now our starting point was the understanding of the discrete problem 
as Tikhonov regularization \cite{A. N. Tikhonov} of the ill-posed (unstable) continuous inverse problem. In Tikhonov
method, regularization is achieved by restricting the set on which solutions are sought and, accordingly, changing the 
criterion for the closeness of the approximate solution to the exact one (by introducing a regularizing operator). In our 
approach, we explicitly introduce the lattice spacing as the regularization parameter, and by varying the lattice spacing — that is, 
by changing the form of the functional to be minimized that defines the criterion for the closeness of the solution to the 
exact one — we carry out the transition to the continuum limit, each time remaining within the class of stable solutions; this is 
exactly what guarantees the existence of such a limit, see below. This idea indeed proved fruitful: we found explicit conditions 
for the existence of the continuum limit. We also found that following the regularization algorithms leads not to a limit for the 
potential matrix itself (such a limit indeed does not exist), but rather for a certain combination of the matrix elements of the 
potential, and this limit is strictly the desired potential in the continuous case.

We shall present the material as follows. First, we derive the discrete inverse problem for the Schrödinger operator on a finite 
interval within the Gelfand–Levitan approach, following the already established version in \cite{Ch}. In doing so, we shall 
partially reproduce the results published there, or present them in a revised form adapted to our problem. This is necessary 
to ensure the coherence of the presentation and to avoid logical gaps that would inevitably arise if such a derivation was omitted. 
We then proceed directly to deriving the inverse problem equations in the continuum limit, following the logic of the Tikhonov
approach. At the same time, we provide all the necessary methodological support while avoiding unduly prolonging the 
presentation or repeating well-known results.

\section{Discrete version of inverse problem on finite interval}

Let us write down the following finite-difference Schr\"odinger equation in the
discrete variable $x_{n}, n \in {\mathbb Z}$ with the step $\Delta$:
\begin{eqnarray}
-\frac{\Psi(x_{n+1},E)-2 \Psi(x_{n},E)+\Psi(x_{n-1},E)}{\Delta ^2}+
W(x_{n}) \Psi(x_{n},E) + \nonumber \\ u(x_{n}) \Psi(x_{n+1},E) +
u(x_{n-1}) \Psi(x_{n-1},E) = E \Psi(x_{n},E),
\label{fdsch}
\end{eqnarray}
$W(x_{n})$ and $u(x_{n})$  being the real values. Here, we intentionally
single out the terms corresponding to the kinetic energy operator (the finite-difference
second derivative).
Note the terms $u(x_{n})$ correspond to a
"minimally non-local" interaction.

Let us consider the bounded interval $[0, \pi]$ with finite number
N of points inside: $x_{0}=0; \enskip x_{N+1}=\pi$, so that
$\Delta = x_{n+1}-x_{n}=\pi/(N+1)$.  Let us supplement the Eq. (1) with the
following (Dirichlet) boundary conditions:
\begin{equation}
\Psi(x_{0},E)=\Psi(x_{N+1},E)=0. \label{fdbc}
\end{equation}

The Sturm-Liouville problem (\ref{fdsch}), (\ref{fdbc}) can be rewritten
in a more visible form by using the symmetric tridiagonal (Jacobi) matrix
($N \times N$)
\begin{eqnarray*}
{\hat H} =  {\hat T} + {\hat V},
\end{eqnarray*}
\begin{eqnarray*}
{\hat T} = \left( \begin{array}{ccccccc} 2/\Delta ^{2} &
-1/\Delta ^2 & 0 & 0 & .  & .  & .\\ -1/\Delta ^2
& 2/\Delta ^{2} & -1/\Delta ^2 & 0
& .  & .  & . \\ 0 & -1/\Delta ^2 & 2/\Delta ^{2}
& -1/\Delta ^2 & . & . & .\\ .  & .  & . & . & . & . & .\\ .
& . & . & .  & . & . & .\\ . & . & . & . & 0 & -1/\Delta ^2
& 2/\Delta ^{2} \end{array} \right),
\end{eqnarray*}
\begin{equation}
{\hat V}= \left( \begin{array}{ccccccc}  W(x_{1}) &
u(x_{1}) & 0 & 0 & .  & .  & .\\ u(x_{1})
& W(x_{2}) & u(x_{2}) & 0
& .  & .  & . \\ 0 & u(x_{2}) & W(x_{3})
& u(x_{3}) & . & . & .\\ .  & .  & . & . & . & . & .\\ .
& . & . & .  & . & . & .\\ . & . & . & . & 0 & u(x_{N-1})
& W(x_{N}) \end{array} \right). \label{HJ}
\end{equation}
The spectrum of the problem (\ref{fdsch}), (\ref{fdbc}) is a set of discrete energy
levels $\{E_{\nu
}\}_{\nu=1}^{N}$ (spectrum points) for bound states  $\Psi _{\nu } \in {\mathbb
R}^{N}$:
\begin{eqnarray*} {\hat H} \Psi _{\nu }= E_{\nu } \Psi _{\nu }, \nonumber
\end{eqnarray*}
\begin{eqnarray*}
\Psi _{\nu }\equiv \left( \begin{array}{c} \Psi (x_{1},E_{\nu }) \\
\Psi (x_{2},E_{\nu }) \\
. \\
. \\
. \\
\Psi (x_{N},E_{\nu })
\end{array} \right),
\end{eqnarray*}
$$\sum_{n=0}^{N+1}
\Delta \Psi_{\mu }(x_{n}) \Psi_{\nu }(x_{n}) = \delta_{\mu \nu}.$$

Besides the energy levels, let us introduce additional fundamental
spectral parameters, normalizing constants or spectral weight factors.
By the definition, these are the coefficients $c_{\nu }$ of
proportionality between the normalized eigenstates $\Psi_{\nu }(x_{n})$ and
the regular solutions $\varphi(x_{n},E_{\nu })$ at the eigenvalue energy,
$\varphi(0,E)=0, \varphi(x_{1},E)=\Delta$ [i.e. the derivative is equal to
1], $\varphi_{\nu }(x_{n}) \equiv \varphi(x_{n},E_{\nu}$):
\begin{eqnarray} \Psi_{\nu }(x_{n})=c_{\nu }\varphi_{\nu }(x_{n}).
\label{pphifd} \end{eqnarray}

The corresponding Parseval's completeness relation is
\begin{eqnarray} \sum_{\nu =1}^{N} \Psi _{\nu }(x_{m})
\Psi _{\nu }(x_{n}) = \delta_{mn}/\Delta \label{Pars}. \end{eqnarray} 
It can
be rewritten  using Eq.(\ref{pphifd}) as follows
\begin{eqnarray} \sum_{\nu
=1}^{N} c_{\nu }^{2} \varphi_{\nu }(x_{m}) \varphi_{\nu
}(x_{n})=\delta_{mn}/\Delta.  \label{pars1} \end{eqnarray}

There are $N$ levels
$E_{\nu}$ and $N-1$ independent parameters $c_{\nu}$. The last is true by virtue 
of the relation $\sum_{\nu =1}^{N} c_{\nu }^{2} =1/\Delta^{3}$ that follows from
(\ref{pars1}) for $n=m=1$. So we have $2 N-1$ spectral parameters, and this number 
is equal to number of interaction parameters : $N$ coefficients $W(x_{n})$ and $N-1$ 
parameters $u(x_{n})$.  Hence, once we know all $E_{\nu }$'s and
the first eigenvector components --- by virtue of Eq.  (\ref{pphifd}) these
are $\Delta c_{\nu }$ in our problem - we can await unambiguous reconstruction of 1-band, i.e.,
three-diagonal Hamiltonian (\ref{HJ}), see e.g. \cite{Ges}. In a general case, a symmetric 
p-band matrix (a matrix with $2 p+1$
bands, p bands below the diagonal) may be uniquely constructed (apart from
certain sign ambiguities) from its eigenvalues and the first p components
of its normalized eigenvectors \cite{GladWil}. 
That is, by solving the problem (\ref{fdsch}), (\ref{fdbc}), we construct the mapping
\begin{eqnarray}
A  {\hat V} = {\hat S}, 
 \label{DirectP} \end{eqnarray} 
where $ {\hat S}$ means the set $\{E_{\nu} , c_{\nu}\}$. This is what is called the "direct problem". Whereas 
recovering ${\hat V}$ from $ {\hat S}$ is the inverse problem. 
The relations (\ref{pars1}) can be viewed as orthogonality relations, but the components of the 
vectors $\varphi_{\nu }(x_{n})$ are now indexed by the energy indices $\nu$, while $x_{n}$ denote 
the indices of the basis vectors. The inner product is defined through the components of the 
normalization vectors $c_{\nu}$. If we change the spectral parameters 
$\{\stackrel{\circ}{E}_{\nu}, \stackrel{\circ}{c}_{\nu} \}\to\{E_{\nu} , c_{\nu}\}$, 
which will also imply the change $\hat{\stackrel{\circ}{V}}\to\hat{V}$, the orthogonality 
relation will be violated, and, basically,  the inverse problem procedure itself will, one way or another, lead  
to restoring this relation. We do not yet know the corresponding algorithms, but we can carry out the 
reorthogonalization simply by using standard methods of linear algebra. Once we know $\hat{\stackrel{\circ}{V}}$ 
and, hence, $\{\stackrel{\circ}{E}_{\nu}, \stackrel{\circ}{c}_{\nu} \}$  from the direct problem (\ref{DirectP}) , we thereby
carry out  the restoration of $\varphi_{\nu }(x_{n})$ to the unknown $ {\hat V}$ , i.e. solve the inverse problem. 

We will use the Gram–Schmidt algorithm; for $N$ vectors, this procedure is defined by $2N - 1$ independent 
parameters (we need also reorthonoramalization), which corresponds to the number of independent spectral parameters 
in our problem. If we used a different orthogonalization scheme, we would have to rewrite our problem so that the number 
of spectral parameters matched the number of parameters of the reorthogonalization algorithm, i.e., we would need to 
introduce the first $p > 1$ components of the eigenvectors. For Gram-Schmidt reorthogonalization, we shall use, 
as a first unit vector, the $\stackrel{\circ}{\varphi}_{\nu}(x_{1})$

Performing these manipulations leads to the following (regular) solutions to the new Schr\"odinger equation
for the spectral set $\{E_{\nu} , c_{\nu}\}$ (see \cite{Ch} for more detail):
\begin{eqnarray} \varphi (x_{m},E_{\nu})= \stackrel{\circ}{\varphi
}(x_{m},E_{\nu}) + \sum_{n=1}^{m-1} \Delta K(x_{m},x_{n})
\stackrel{\circ}{\varphi }(x_{n},E_{\nu}), \label{fKfnm} \end{eqnarray}
where the coefficients K [the kernel of the transformation operator
(\ref{fKfnm})] follow from the conditions of the orthogonality of
new vectors  (by measure $c_{\nu}^2$) $\varphi (x_{m},E_{\nu }) \,
(m=1,2,...N)$:  $$\varphi (x_{m
> n},E_{\nu}) \perp \varphi (x_{n},E_{\nu}),$$ which lead
to the system  of algebraic equations for K -- discrete analog
of central equations of the inverse problem:
\begin{eqnarray} K(x_{m},x_{n})+Q(x_{m},x_{n})+ \sum_{p=1}^{m-1} \Delta
K(x_{m},x_{p}) Q(x_{p},x_{n})=0, \quad m>n, \label{GLnm}
\end{eqnarray}
where $Q(x_{n},x_{m})$ is determined by
\begin{eqnarray}
Q(x_{m},x_{n}) = \sum_{\nu =1}^{N} c_{\nu }^{2} \stackrel{\circ}{\varphi
}(x_{m},E_{\nu}) \stackrel{\circ}{\varphi }(x_{n},E_{\nu})  \nonumber \\
- \sum_{\mu =1}^{N}\stackrel{\circ}{c}_{\mu }^{2}
\stackrel{\circ}{\varphi }(x_{m},\stackrel{\circ}{E}_{\mu})
\stackrel{\circ}{\varphi }(x_{n}, \stackrel{\circ}{E}_{\mu}).
\label{QN} \end{eqnarray}

It should be noted that the formula (\ref{fKfnm}) is valid
for any energy values $E$  \cite{Ch}:
\begin{eqnarray} \varphi (x_{m},E)= \stackrel{\circ}{\varphi }(x_{m},E) +
\sum_{n=1}^{m-1} \Delta K(x_{m},x_{n}) \stackrel{\circ}{\varphi }(x_{n},E).
\label{fearb1} \end{eqnarray}
Let us stress here that $K$ is independent of energy $E$.
The formulas (\ref{GLnm}), (\ref{QN}) and (\ref{fearb1})
give the expression for $K$ in the form of sum of
products of the old solutions and transformed ones:
\begin{eqnarray}
K(x_{m},x_{n}) = -\sum_{\nu }^{N} c_{\nu}^{2}\varphi
(x_{m},E_{\nu}) \stackrel{\circ}{\varphi }(x_{n},E_{\nu}) +
\sum_{\mu }^{N}\stackrel{\circ}{c}^{2}_{\mu } \varphi
(x_{m},\stackrel{\circ}{E}_{\mu}) \stackrel{\circ}{\varphi
}(x_{n},\stackrel{\circ}{E}_{\mu}). \nonumber \\
\label{kp0pnm} \end{eqnarray}

It remains to obtain equations for the transformed potentials $V$
and $u$. We already know the solutions of Eq. (\ref{fdsch}) with
the unknown potentials $W(x_{n})$ and $u(x_{n})$ [see formulas
(\ref{fKfnm}) and (\ref{GLnm})], i.e. eigenvectors of the new
Hamiltonian plus associated eigenvalues $E_{\nu }$.  Let us
multiply both parts of the Schr\"odinger equation (\ref{fdsch})
for the solutions $\varphi (x_{m},\stackrel{\circ}{E}_{\mu})$  [Eq.
(\ref{fearb1})] and $\stackrel{\circ}{\varphi} (x_{n},\stackrel{\circ}{E}_{\mu})$  by $\stackrel{\circ}{\varphi
}(x_{n}, \stackrel{\circ}{E}_{\mu})$ and
$\varphi(x_{m},\stackrel{\circ}{E}_{\mu})$, respectively, sum over
$\mu $ with weight $\stackrel{\circ}{c}_{\mu }^{2}$ and subtract
from each other the resulting expressions. At fixed $m$, we
perform this procedure for $n=m, m-1,...$.  In calculating sums
(over $\mu$) one should take into account the relation
(\ref{pars1}).  As a result, we get to the following equations for
$W$ and $u$:  \begin{eqnarray} \{W(x_{m}) -
\stackrel{\circ}{W}(x_{n})\} K(x_{m},x_{n})+ u(x_{m})
K(x_{m+1},x_{n}) - \stackrel{\circ}{u}(x_{n}) \nonumber \\ \times
K(x_{m},x_{n+1}) + u(x_{m-1})
K(x_{m-1},x_{n})-\stackrel{\circ}{u}( x_{n-1}) K(x_{m},x_{n-1})
\nonumber \\ = \frac{K(x_{m+1},x_{n})-2
K(x_{m},x_{n})+K(x_{m-1},x_{n})}{\Delta^2} \nonumber \\ -
\frac{K(x_{m},x_{n+1})-2
K(x_{m},x_{n})+K(x_{m},x_{n-1})}{\Delta^2}, \enskip n \le m-2;
\label{vfd1} \end{eqnarray} and for $n=m,m-1$
\begin{eqnarray} \left\{ \begin{array}{l}
\frac{u(x_{m-1})-\stackrel{\circ}{u}(x_{m-1})}{\Delta} =
\frac{K(x_{m+1},x_{m-1})-K(x_{m},x_{m-2})}{\Delta ^2} \\
- W(x_{m}) K(x_{m},x_{m-1})+\stackrel{\circ}{W}(x_{m-1}) K(x_{m},x_{m-1}) \\
- u(x_{m}) K(x_{m+1},x_{m-1})+\stackrel{\circ}{u}(x_{m-2}) K(x_{m},x_{m-2}),
\enskip n=m-1 \\
\frac{W(x_{m})-\stackrel{\circ}{W}(x_{m})}{\Delta}=
\frac{K(x_{m+1},x_{m})-K(x_{m},x_{m-1})}{\Delta ^2} \\
- u(x_{m}) K(x_{m+1},x_{m}) + K(x_{m},x_{m-1}) \stackrel{\circ}{u}
(x_{m-1}), \enskip n=m,
\end{array}
\right.
\label{vfd2}
\end{eqnarray}
and $u(x_{m})=\stackrel{\circ}{u}(x_{m}), \enskip  n=m+1$.  

The fact that we obtained the unchanged coefficient $u_{m}$ is not a contradiction, since 
we reconstructed only a part of the potential ("unfinished" reconstruction) from the 
first $n+1$ reorthogonalized unit vectors. 
On the other hand, if we replace $-1/\Delta ^2 + \stackrel{\circ}{u}(x_{m})$ by $c [-1/\Delta ^2 + \stackrel{\circ}{u}(x_{m})]$ 
for any $c$, and  $\stackrel{\circ}{\varphi}(x_{m+1},\stackrel{\circ}{E}_{\mu})$ 
by $\tilde{\stackrel{\circ}{\varphi}}(x_{m+1},\stackrel{\circ}{E}_{\mu})=\stackrel{\circ}{\varphi}(x_{m+1},\stackrel{\circ}{E}_{\mu})/c$
this will not change the Schrödinger equation (\ref{fdsch}) for  $\stackrel{\circ}{\varphi}(x_{m},\stackrel{\circ}{E}_{\mu})$  and if we 
carry out the above procedure but instead of $\stackrel{\circ}{\varphi}(x_{m+1},\stackrel{\circ}{E}_{\mu})$  use 
 $\tilde{\stackrel{\circ}{\varphi}}(x_{m+1},\stackrel{\circ}{E}_{\mu})$ (for $\varphi(x_{m+1}, E)$ the form (\ref{fKfnm}) is kept) 
we will obtain 
\begin{eqnarray}
u(x_{m})=(1-c) /\Delta ^2 + c\stackrel{\circ}{u}(x_{m}).
\label{uxm}
\end{eqnarray}
So, we can see that in fact there is no unique expression for $u(x_{m})$, but at the same time this 
allows us to uniquely reconstruct the entire matrix. Indeed, suppose we know in advance 
the value for the reconstructed $u(x_{m})$. This assertion by no means contradicts to
(\ref{uxm}) because we can always set $c$ such that $u(x_{m})$ just 
coincides with that value for any $\stackrel{\circ}{u}(x_{m})$.  From
(\ref{vfd2}) we obtain $W(x_{m})$ and $u(x_{m-1})$.
We then substitute the value $u(x_{m-1})$ (by virtue of the
symmetry of potential matrix) into equations (\ref{vfd1})  or (\ref{vfd2}) for
$\bar{m}=m-1$, apply  the manipulations (\ref{uxm}) again, from which we find, in turn, $V(x_{m-1})$ and
$u(x_{m-2})$ and so on. From  (\ref{uxm}) we can see that for $c \ne 1$ the limit $\Delta \to 0$ does not exist, 
this alone indicates the impossibility of taking the limit for the potential matrix. The only exception 
is $c=1$. We will soon see that it is precisely in this case that everything will work out for us. But the question remains about 
defining $u$ at the edge of our interval $x_{N}$. At that point  the situation is different -
 we no longer have an orthogonalized vector 
at the nearby point $x_{N+1}$. But we can use the Schrödinger equation to obtain information about the regular solution at the point $x_{N+1}$.
So, instead of  (\ref{vfd2}), we have the following system of
equations
\begin{eqnarray} \left\{ \begin{array}{l}
\frac{u(x_{N-1})-\stackrel{\circ}{u}(x_{N-1})}{\Delta} =
\frac{\sum_{\mu =1}^{N} \stackrel{\circ}{c}_{\mu }^{2} \varphi
(x_{N+1},\stackrel{\circ}{E}_{\mu}) \stackrel{\circ}{\varphi
}(x_{N-1},\stackrel{\circ}{E}_{\mu})-K(x_{N},x_{N-2})}{\Delta ^2} \\
- W(x_{N}) K(x_{N},x_{N-1})+\stackrel{\circ}{W}(x_{N-1}) K(x_{N},x_{N-1}) \\
- u(x_{N}) \sum_{\mu =1}^{N} \stackrel{\circ}{c}_{\mu }^{2}
\varphi (x_{N+1},\stackrel{\circ}{E}_{\mu})
\stackrel{\circ}{\varphi
}(x_{N-1},\stackrel{\circ}{E}_{\mu})+\stackrel{\circ}{u}(x_{N-2})
K(x_{N},x_{N-2}), \\
\frac{W(x_{N})-\stackrel{\circ}{W}(x_{N})}{\Delta}=
\frac{\sum_{\mu =1}^{N} \stackrel{\circ}{c}_{\mu }^{2} \varphi
(x_{N+1},\stackrel{\circ}{E}_{\mu}) \stackrel{\circ}{\varphi
}(x_{N},\stackrel{\circ}{E}_{\mu})-K(x_{N},x_{N-1})}{\Delta ^2} \\
- u(x_{N}) \sum_{\mu =1}^{N} \stackrel{\circ}{c}_{\mu }^{2}
\varphi (x_{N+1},\stackrel{\circ}{E}_{\mu})
\stackrel{\circ}{\varphi }(x_{N},\stackrel{\circ}{E}_{\mu}) +
K(x_{N},x_{N-1}) \stackrel{\circ}{u} (x_{N-1}),
\end{array}
\right. \nonumber \\ \label{vfdN}
\end{eqnarray}
where $\varphi (x_{N+1},\stackrel{\circ}{E}_{\mu})$ obeys
the Schr\"odinger equation (\ref{fdsch}).
We can see that $\varphi
(x_{N+1})$  falls out of our derivation as direct substitution of  $\varphi
(x_{N+1},,\stackrel{\circ}{E}_{\mu})$ from (\ref{fdsch}) to (\ref{vfdN}) gives identity. 
Of course, the obvious and direct way is deriving $u(_{N-1})$ and $W(x_{N})$ from  the boundary conditions 
($\varphi(x_{N+1},E_{\nu})=0$ in our case). 
Nevertheless, the system (\ref{vfdN}) is useful for the proof of the uniqueness of such an algorithm. In fact, 
we can consider ${\hat V}$ as an initial system whereas $\hat{\stackrel{\circ}{V}}$ is reffered to the transformed 
one, $\hat{V}  \rightarrow \hat{\stackrel{\circ}{V}}$, $\{E_{\nu} , c_{\nu}\} \rightarrow \{\stackrel{\circ}{E}_{\mu} , \stackrel{\circ}{c}_{\mu}\}$ and 
$\varphi  \rightarrow \stackrel{\circ}{\varphi}$.
By virtue of (\ref{kp0pnm}), such a replacement leads to $K(x_{m},x_{n}) \rightarrow -K(x_{n},x_{m})$. Now we can  
carry out the transition  $\hat{V}  \rightarrow \hat{\stackrel{\circ}{V}}$ and respectively rewrite Eqs. (\ref{vfdN}) taking these replacements into account 
with new equations being expressed in terms of the known functions.

\section{Continuum limit}

Let us now pass to the limit of the continuum variable $x$, i.e.
to the limit $\Delta \to 0$ ($N \to \infty$) so that $\Delta N =
\pi N/(N+1) \to \pi$ in the formulas (\ref{fKfnm}), (\ref{GLnm}),
(\ref{vfd1}) and (\ref{vfd2}).  We can formally write down the expression for 
the kernel $K$ of the operator (\ref{phitr}) which transforms the
solutions  to the initial potential into the solutions to the new
one as follows:
\begin{eqnarray} K(x,y)+Q(x,y)+\int_{0}^{x} K(x,z)Q(z,y)dz=0,
\label{kq} \end{eqnarray} where the kernel Q is constructed from
the unperturbed functions with the old and new spectral parameters
[as in Eq.  (\ref{QN})]:  \begin{eqnarray} Q(x,y)=\sum_{\nu
}c_{\nu}^{2}\stackrel{\circ}{\varphi }(x,E_{\nu})
\stackrel{\circ}{\varphi }(y,E_{\nu})- \sum_{\mu
}\stackrel{\circ}{c}^{2}_{\mu } \stackrel{\circ}{\varphi
}(x,\stackrel{\circ}{E}_{\mu}) \stackrel{\circ}{\varphi
}(y,\stackrel{\circ}{E}_{\mu}).  \label{Q3}
\end{eqnarray}

In the continuum case, we have $\varphi(0,E)=0, \enskip
\varphi'(0,E)=1$. Norming parameters are in that case, too,
the coefficients of proportionality between normalized
eigenfunctions and regular solutions. The multiplication by
$c_{\nu} = 1/\int_{0}^{\pi} \varphi^{2}(x,E_{\nu})dx$ turns
regular solution (at $E=E_{\nu}$) into the normalized one,
$$\Psi'(x,E_{\nu})|_{x=0} =c_{\nu}.$$
The continuum analog of the expression for the transformed regular solutions has the
following form 
\begin{eqnarray} \varphi(x,E_{\nu})=\stackrel{\circ}{\varphi
}(x,E_{\nu})+ \int_{0}^{x} K(x,y)\stackrel{\circ}{\varphi }(y,E_{\nu})dy,
\label{phitr}
\end{eqnarray}
and similarly
\begin{eqnarray} \varphi(x,E)=\stackrel{\circ}{\varphi
}(x,E)+ \int_{0}^{x} K(x,y)\stackrel{\circ}{\varphi }(y,E)dy,
\label{phitr1}
\end{eqnarray}
 where $x \in [0,\pi]$.

The fundamental problem here is that the mapping (\ref{DirectP}), as is known, becomes an ill-posed 
problem in the continuum case. That is, the slightest changes in $S$ correspond to finite $\delta V$. 
Without resolving the issue of the stability of the problem at each step of the approximation $x \to 0$, 
we have no reason to expect the discrete problems (\ref{DirectP}) to converge to the continuum analogue.
The Tikhonov method makes it possible to solve this problem as follows. The concept of an approximate 
solution $V_{\cal{A}} \in {\mathbb V}$ of equation (\ref{DirectP}) is introduced, such that $V_{\cal{A}}$ minimizes the functional
$M_{\Delta}[S_{\cal{A}}, V]=\rho_{{\mathbb S}}(A V, S_{\cal{A}})+\Delta \Omega[V]$, where $\rho_{{\mathbb S}}$ is the metric
in the space ${\mathbb S} $. The functional $\Omega[V]$ is called a stabilizer of the problem $A V =S$.
To make it clear, the exact solution $V_{\cal{E}}$ minimizes the functional 
$\rho_{{\mathbb S}}(A V_{\cal{E}}, S_{\cal{E}})$ for exact value $S_{\cal{E}}$ , i.e. it corresponds to $\rho_{{\mathbb S}}(A V_{\cal{E}}, S_{\cal{E}})=0$, 
but it is unstable. Therefore, we deliberately eliminate  "dangerous" points $S$ by restricting the space of admissible solutions, i.e. 
we consider the problem of finding a constrained extremum, when we work with a modified functional and introduce the Lagrange multiplier 
$\Delta$. The finite-difference approach with the finite step $\Delta$ exactly implements such a scheme. In this case, 
the functional $M_{\Delta}[S_{\cal{A}}, V]$ is a distance (metric) between points in the finite-dimensional space 
${\bf S} \subseteq {\mathbb R} ^{2N-1}$.  The method must 
guarantee that as the approximation $S_{\cal{A}}$ approaches the exact value $S_{\cal{E}}$ as ${\Delta \to 0}$, the approximate solution  $V_{\cal{A}}$
would tend to the desired exact solution $V_{\cal{E}}$ of the equation (\ref{DirectP}) . Explicitly, we need to construct an algorithm such that 
the following property holds: there exists a function $\Delta (\delta)$ such
that for any $\varepsilon > 0$ one can find the number $\delta(\varepsilon)$ such that if 
$\rho_{{\mathbb S}}(\{E_{\nu}, c_{\nu}\}_{\cal{E}}, \{E_{\nu}, c_{\nu}\}_{\cal{E}}
+\delta \{E_{\nu}, c_{\nu}\}) < \delta(\varepsilon)$ then $\rho_{\mathbb V}(V_{\cal{E}}, V_{\cal{A}})< \varepsilon$, 
where $\rho_{\mathbb V}$ is the metric in the space ${\mathbb V}$.

The implementation of such a scheme is carried out within the procedure described by the equations (\ref{vfd1}) and (\ref{vfd2}), 
with  the latter ones providing a new context for our problem. Indeed, the equations determine $W$ and $u$, while at the same time 
they can also be viewed as equations for the coefficients $K$. We will make use of this circumstance. We must establish an obvious 
requirement that $K$ and its derivatives converge to the continuum limit at any point. To this end, consider 
a convergent sequence ${\cal X}_{x}$, such that  for the partition $N$ 
$$\{{\cal X}_{x}\}_{N}=z: (x-z)=\min_{\stackrel{0<n<N+1}{x-x_{n}>0}}(x-x_{n})$$ 
for any $x \in [0,\pi]$. 
We can also define sequences for the corresponding values of the functions and their derivatives, for example for the
second (finite-difference) derivative $(D^2f)_{x}$ in (\ref{fdsch}). 
We will choose $S'_{\cal{A}}$ and $S''_{\cal{A}}$, $\rho_{{\mathbb S}}(S'_{\cal{A}}, S''_{\cal{A}})<\delta(\varepsilon)$, and corresponding partitions $N'$ and $N''$  (by using (\ref{GLnm}, \ref{QN})) so that to 
construct a fundamental sequence for $(D^{2}_{x}K)_{x,y}$  (and likewise for $(D^{2}_{y}K)_{x,y}$):
\begin{eqnarray}
\{(D^{2}_{x}K)_{x,y}\}_{N'}-\{(D^{2}_{x}K)_{x,y}\}_{N''}<\varepsilon; \enskip N', N'' >{\bar N}
\label{fundseq}
\end{eqnarray}
for any $(x,y) \in [0,\pi]$, $\varepsilon>0$ and some number ${\bar N}(\varepsilon)$. The condition (\ref{fundseq}) 
is equivalent to the fact that the sequences $(D^{2}_{x}K)_{x,y}$, $(D^{2}_{y}K)_{x,y}$ and, similarly, $K_{x,y}$ tend 
to $\frac{\partial ^2}{\partial x^2}K(x,y)$,   $\frac{\partial ^2}{\partial y^2}K(x,y)$ and $K(x,y)$, respectively, 
for any $(x,y)$. Moreover, it follows from Eqs. (\ref{vfd1}) and (\ref{vfd2}) that the potential coefficients $W$ and $u$ also have limits, let us call 
them $W(x)$ and $u(x)$. Here lies a fundamental feature of our problem - the existence of these limits does not imply the existence of 
some continuous functions $W(x)$ and $u(x)$, defined for all $x$. The contrary would mean that we had come to the limit of the matrix $\hat {V}$, 
which as we now know is impossible. Instead, as we will now see, the discussion will concern the existence of a certain combination of $W(x)$ 
and $u(x)$ as a continuous function.

Such a combination arises immediately when considering the limit for equations (\ref{vfd1}), which, as is easy to see, has the form 
\begin{eqnarray}
\{W_{d}(x)-
\stackrel{\circ}{W}_{d}(y) + u_{d}(x)- \stackrel{\circ}{u}_{d}(y)\}
K(x,y) = \frac{\partial ^2}{\partial x^2}K(x,y) -
\frac{\partial ^2}{\partial y^2}K(x,y), \enskip x>y, \label{vdel20}
\end{eqnarray}
where  $W_{d}(x) \equiv \lim_{N \to \infty} \{W_{x}\}_{N}$ and, analogously, $u_{d}(x) \equiv
\lim_{N \to \infty} \{u_{x}\}_{N}$, and we used $c=1$ (see above).
The transition to the continuum limit in Eqs. (\ref{vfd2}) is in fact a nontrivial 
task. As a first step, let us write down the result of the limiting transition for the finite-difference terms - this result has already 
been obtained in the paper \cite{Ch}, and here we will not repeat the derivation: 
\begin{eqnarray}
\frac{K(x_{m+1},x_{m-1})-K(x_{m},x_{m-2})}{\Delta ^2}
\longrightarrow \frac{\partial ^2}{\partial x^2}K(x,y)|_{y=x} -
\frac{\partial ^2}{\partial y^2}K(x,y)|_{y=x} \nonumber \\ +
\Delta ^{-1} [\frac{d}{dx}K(x,x)+O[\Delta)], \label{lim1}
\end{eqnarray}
\begin{eqnarray}
\frac{K(x_{m+1},x_{m})-K(x_{m},x_{m-1})}{\Delta ^2}  \longrightarrow  \Delta ^{-1}[ \frac{d}{dx}K(x,x)+O[\Delta)],
\enskip \Delta \rightarrow 0. \label{lim2}
\end{eqnarray}
The "imbalance" of these expressions immediately stands out - the presence of terms differing by order of 
magnitude $\Delta^{-1}$, the parameter of our theory, which immediately suggests that the problem is unstable. 
The only way out is that the entire sum of the non-diverging terms vanishes, which, as we will now see, 
is the case. Indeed, summing Eqs. (\ref{vfd2}), assuming $c=1$ (i.e. $u(x_{m})=\stackrel{\circ}{u}(x_{m})$), and, finally, 
collecting the terms with factor $\Delta^{-1}$, we obtain
\begin{eqnarray} 
R(x) \equiv -2\frac{d}{dx}K(x,x)+W_{d}(x) -  \stackrel{\circ}{W}_{d}(x) + u_{d}(x)
-\stackrel{\circ}{u}_{d}(x) = 0.
\label{R}
\end{eqnarray}
Let us go further along the path of resolving the uncertainty $R(x)/\Delta$:
\begin{eqnarray} 
\lim_{\Delta \to 0} \frac{R(x)}{\Delta}=\lim_{\Delta \to 0} \frac{d R(x)}{d\Delta}= 
\lim_{\Delta \to 0}[ \frac{d[-2\frac{d}{dx}K(x,x)]}{dV}\frac{dV}{d \Delta} ]
+\lim_{\Delta \to 0}\frac{dV(x)}{d \Delta}, \nonumber \\ 
V(x) \equiv W_{d}(x)  + u_{d}(x) + \stackrel{\circ}{u}_{d}(x).
\label{lim}
\end{eqnarray}
Employing  (\ref{R}) gives the limit (\ref{lim}) is zero strictly. But then the combination of all non-diverging terms in the sum of Eqs. (\ref{vfd2}) 
tends to zero (we assume $ \stackrel{\circ}{u}(x_{m-1})= \stackrel{\circ}{u}(x_{m})$ in 
the continuum limit, which is always possible):
\begin{eqnarray} 
\frac{\partial ^2}{\partial x^2}K(x,y)|_{y=x}
- \frac{\partial ^2}{\partial y^2}K(x,y)|_{y=x} -\{W_{d}(x) -
\stackrel{\circ}{W}_{d}(x) +  u_{d}(x)  \nonumber \\
-\stackrel{\circ}{u}_{d}(x)\} K(x,x) =0.  
\label{lim3}
\end{eqnarray}
We would like to specially emphasize that the application here of the technique of differentiating 
a composite function in (\ref{lim}) became 
possible precisely because of maintenance of the connection between $\Delta$ and $S_{\cal A}$, see (\ref{fundseq}),
in the Tikhonov method described above.

The Eqs. (\ref{vdel20}), (\ref{R}) and (\ref{lim2}) can now be rewritten as
\begin{eqnarray} \left\{ \begin{array}{l}
\{V(x)- \stackrel{\circ}{V}(y)\} K(x,y)  \\ =
\frac{\partial ^2}{\partial x^2}K(x,y) -
\frac{\partial ^2}{\partial y^2}K(x,y)
\\
V(x)-\stackrel{\circ}{V}(x) = 2 \frac{d}{dx}K(x,x) \\
\end{array}
\label{vdel22}
\right.
\end{eqnarray}
This system (added by $K(0,0)=0$) represents the classical Goursat
problem (hyperbolic-type equation with boundary conditions on a characteristic) and its solvability follows
from the well known theorems. These equations (\ref{vdel22}) represent the continuum limit for 
the operator (\ref{DirectP}) and, hence, we have now proved the existence inverse eigenvalue problem as 
derived from approximate discrete version.  Note that no other combinations in (\ref{vfd2}), nor the 
equations (\ref{vfd2}) themselves individually, could
lead to the desired result.  For example, if we only use the first equation in (\ref{vfd2}), we will not be able 
to consider $K(x,y)$ as a function of some $V(x)$. Such subtle relationships may represent an obstacle to
deriving continuum equations based on the general theory of p-band matrix.  The symmetry in equations 
(\ref{vdel22}) $V  \leftrightarrow \stackrel{\circ}{V}$
allows us to state unambiguously that the function $V(x)$ is indeed the desired potential.

\section{Conclusions}
In the present paper we carried out the derivation of main
formulas of the inverse eigenvalue problem for the Sturm-Liouville operator 
on the base of its discrete approximation. At the same time, we sticked to Tikhonov concept, 
where the discrete approximation is performed as a regularization of the ill-posed operator problem 
(\ref{DirectP}). In this approach we 
consistently vary the step of finite-difference differentiation $\Delta$ and the estimate precision
$\delta \{E_{\nu}, c_{\nu}\} $. Following 
this logic  leads to new relations, which makes such a method quite
interesting in itself. For example, we can also consider other regularization options
by applying exactly solvable models, in particular those obtained 
within the Darboux transformation approach. These transformations (and their subclass, known 
as models of supersymmetric quantum mechanics) make it possible to construct potentials 
under variations,  in a given way, of any finite number of spectral parameters.  Such solutions lie
in a broader class compared with exactly solvable models of the inverse 
problem. It would be interesting to see what equations we arrive at in this case if we again apply 
the Tikhonov approach, analogous to that considered above, while increasing to infinity 
the number of varied spectral parameters.


\begin{thebibliography}{100
}

\bibitem{AgranMar}  Z S Agranovich and V A Marchenko 1963
{\it The Inverse Problem of Scattering Theory}  (London: Gordon and
Breach)

\bibitem{Levit}   B M  Levitan and  I S Sargsjan 1975 {\it Introduction
to Spectral Theory}  (Providence, RY: Amer. Math. Soc.)

\bibitem{Mar}  V A Marchenko 1986 {\it Sturm-Liuoville Operators and
Applications} (Basel: Birkhauser)

\bibitem{ChSab} K Chadan  and P Sabatier  1989 {\it Inverse Problems in
Quantum Scattering Theory} 2nd edn (Heidelberg: Springer)

\bibitem{New} R G Newton  1982 {\it Scattering Theory of Waves and
     Particles} 2nd edn (NY: Springer).

\bibitem{Sab} P S Sabatier 2000 Past and future of inverse problems
{\it J. Math. Phys.}  {\bf 41} 4082-124

\bibitem{AblSeg} Mark J. Ablowitz and Harevey Segur Solitons and the Inverse Scattering Transform.   SIAM, 1981

\bibitem{Geramb} M. Sander and H.V. Von Geramb  1995 ``Nucleon‐nucleon potentials from quantum inversion and Arndt’s phase shifts SP82, SM86, FA91, SM93, SP94'' Published in: AIP Conf.Proc. 334 (1995), 825-828
Contribution to: FB14, 825-828


\bibitem{Fad} L D Faddeev 1976 Inverse problem of quantum scattering theory. II.. J Math Sci 5, 334–396

\bibitem{Barg} V Bargmann 1949 Rev. Mod Phys.  {\bf 21}, 488


\bibitem{Ch1} V  M Chabanov and B N Zakhariev 2001
"Resonance absolute quantum reflection at selected energies",
Phys. Rev. Lett., {\bf 87}, N.16, 160408-(1-4)


\bibitem{Ch2} V M Chabanov and B N Zakhariev  1997 "New situation in quantum
mechanics (wonderful potentials from the inverse problem)",
Inverse Problems, {\bf 13}, R47-R79

\bibitem{Ch3}  V M Chabanov and B. N. Zakhariev 1999
"Coexistence of confinement and propagating waves: a quantum paradox",
Phys. Lett. A {\bf 255}, 123-128


\bibitem{CK} K M Case and M Kac 1973  A discrete version
of the inverse scattering problem {\it J. Math. Phys.} {\bf 14} 594-603


\bibitem{C} K M Case  1973 On discrete inverse scattering problems. II
{\it J. Math. Phys.} {\bf 14} 916-20

\bibitem{Cn} K M Case  1974  Orthogonal polynomials from the viewpoint
of scattering theory {\it J.Math.Phys} {\bf 15} 2166-74

\bibitem{Berry} J G Berryman and R R Green 1978 Discrete inverse scattering
theory and the continuum limit {\it Phys.Lett} {\bf 65A} 13-15

\bibitem{Ch} V M Chabanov  2004 ``Inverse eigenvalue problem for the discrete
three-diagonal Sturm-Liouville operator and the continuum limit''
J.Phys.A: Math. Gen. {\bf 37}, 9139–55

\bibitem{A. N. Tikhonov} A N Tikhonov, “On ill-posed problems of linear algebra and a stable method for their solution,” Dokl. Akad. Nauk SSSR, 163:3 (1965), 591–594


\bibitem{Ges} F Gesztesy  and B Simon  1997 M-functions and inverse spectral
analysis for finite and semi-infinite Jacobi matrices {\it J. d'Anal.Math.}
{\bf 73} 267-297

\bibitem{GladWil} G M L Gladwell and N B Willms 1989 A discrete Gel'fand-
Levitan method for band-matrix inverse eigenvalue problems {\it Inverse
Problems} {\bf 5} 165-179



\end{thebibliography}
\end{document}